# The International X-ray Observatory
## Activity submission in response to the Astro2010 Program Prioritization Panel RFI

Jay Bookbinder
Smithsonian Astrophysical Observatory
1-617-495-7058
jbookbinder@cfa.harvard.edu

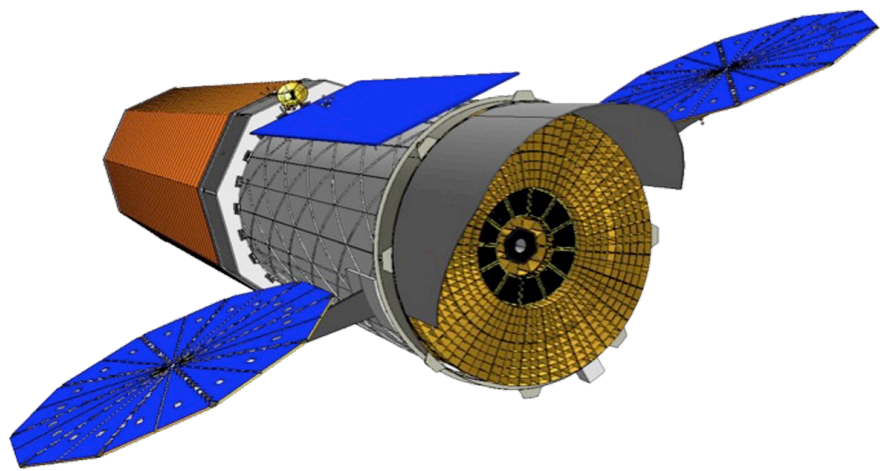

**Submitted on behalf of the IXO Study Coordination Group, whose members are**

Didier Barret (CESR, Toulouse)
Mark Bautz (MIT, Cambridge)
Jay Bookbinder (SAO, Cambridge)
Joel Bregman (Univ. Michigan, Ann Arbor)
Tadayasu Dotani (ISAS/JAXA, Sagamihara) – JAXA Project Manager
Kathryn Flanagan (STScI, Baltimore)
Philippe Gondoin (ESA, Noordwijk) – ESA Study Manager
Jean Grady (GSFC, Greenbelt) – NASA Project Manager
Hideyo Kunieda (Nagoya University, Nagoya) – SCG Co-Chair
Kazuhisa Mitsuda (ISAS/JAXA, Sagamihara)
Kirpal Nandra (Imperial College, London)
Takaya Ohashi (Tokyo Metropolitan University, Tokyo)
Arvind Parmar (ESA, Noordwijk) – ESA Study Scientist, SCG Co-Chair
Luigi Piro (INAF, Rome)
Lothar Strüder (MPE, Garching)
Tadayuki Takahashi (ISAS/JAXA, Sagamihara)
Takeshi Go Tsuru (Kyoto University) – JAXA Study Scientist
Nicholas White (GSFC, Greenbelt) – NASA Project Scientist, SCG Co-Chair

**and on behalf of the 69 members of the IXO Science Definition Team, Instrument Working Group, and Telescope Working Group, whose membership is listed at**
http://ixo.gsfc.nasa.gov/people/

## SUMMARY

The International X-ray Observatory (IXO), a joint NASA-ESA-JAXA effort, will address fundamental and timely questions in astrophysics:

*What happens close to a black hole?*
*How did supermassive black holes grow?*
*How does large scale structure form?*
*What is the connection between these processes?*

To address these science questions, IXO will trace orbits close to the event horizon of black holes, measure black hole spin for several hundred active galactic nuclei (AGN), use spectroscopy to characterize outflows and the environment of AGN during their peak activity, search for supermassive black holes out to redshift z = 10, map bulk motions and turbulence in galaxy clusters, find the missing baryons in the cosmic web using background quasars, and observe the process of cosmic feedback where black holes inject energy on galactic and intergalactic scales.

IXO will employ optics with 20 times more collecting area at 1 keV than any previous X-ray observatory. Focal plane instruments will deliver a 100-fold increase in effective area for high-resolution spectroscopy, deep spectral imaging over a wide field of view, unprecedented polarimetric sensitivity, microsecond spectroscopic timing, and high count rate capability. The improvement of IXO relative to current X-ray missions is equivalent to a transition from the 200 inch Palomar telescope to a 22 m telescope while at the same time shifting from spectral band imaging to an integral field spectrograph.

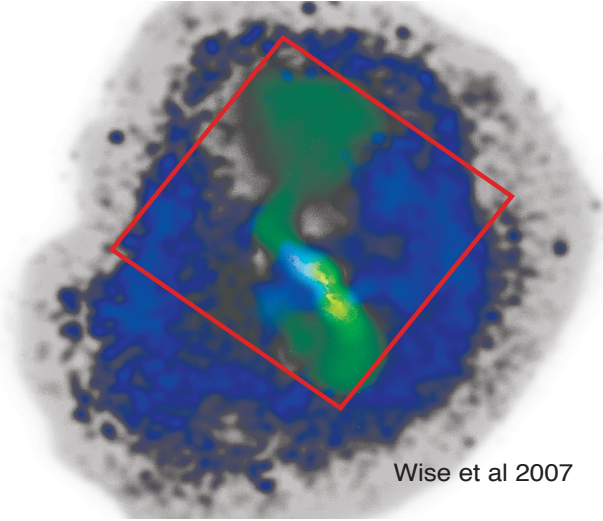

Wise et al 2007

*Supermassive black holes have a profound effect on the growth of structure in the Universe, as shown in this image of Hydra A [Chandra (blue)/VLA (green)]. IXO will study these energetic phenomena via non-dispersive spectral/spatial measurements with high spectral resolution (5 x 5 arcmin XMS FOV overlaid) to determine the temperature, ionization state, and velocities in the intracluster medium.*

The heart of the mission is the X-ray optical system. The 3 $m^2$ collecting area with 5 arcsec angular resolution is achieved using a 20 m focal length deployable optical bench. To reduce risk, two independent optics technologies are under development in the U.S. and in Europe.

The planned instrument complement will employ microcalorimeter arrays (for high-resolution spectroscopic imaging), an active pixel sensor array (for wide-field imaging), CdTe detectors (for hard X-ray imaging), a high-efficiency grating spectrometer, a gas pixel imaging polarimeter, and silicon drift diodes (for high time resolution spectroscopy).

The modular nature of the IXO architecture provides clearly definable interfaces to enable sharing the development between international partners. The IXO spacecraft has substantial redundancy and flight heritage. The mission is on schedule for launch in 2021 to an L2 orbit, with a five-year lifetime and consumables for 10 years.

IXO will be available to the entire astronomical community. Previous experience assures us that unexpected discoveries will abound, and IXO will contribute to the understanding of new phenomena as they are uncovered—a key feature of great observatories (Astro 2010 White Paper: The Value of Observatory-Class Missions, Sembach et al.)

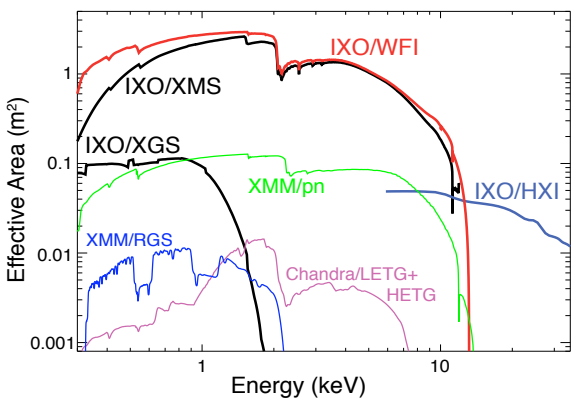

*The IXO effective area will be more than an order of magnitude greater than current imaging X-ray missions. Coupling this with an increase in spectral resolving power up to two orders of magnitude higher relative to previous capabilities will open a vast discovery space for high-energy phenomena.*

# 1. SCIENCE

The extragalactic X-ray sky is dominated by two kinds of sources: accreting supermassive black holes (SMBH) in galactic nuclei, comparable in size to the Solar System, and clusters of galaxies, more than a million light years across. The energy liberated by growing black holes influences the infall of gas in galaxies and clusters, while some analogous process, still poorly understood, ties the growth of black hole mass to a fixed fraction of its host galaxy's bulge.[1,2,3,4] The remarkable link between the two most populous types of extragalactic X-ray sources implies that a two-way connection called "feedback" is a key ingredient of understanding them both.

The driving science goals of IXO are to determine the properties of the extreme environment and evolution of black holes, measure the energetics and dynamics of the hot gas in large cosmic structures, and establish the connection between these two phenomena. IXO will also constrain the equation of state of neutron stars and track the dynamical and compositional evolution of interstellar and intergalactic matter throughout the epoch of galaxy growth. In addition, IXO measurements of virtually every class of astronomical object will return serendipitous discoveries, characteristic of all major advances in astronomical capabilities.

## *Matter Under Extreme Conditions*

Black holes and neutron stars provide the strongest gravitational fields and among the most extreme environments in the Universe. IXO's capabilities will allow us to answer the questions: What are the effects of strong gravity close to a black hole event horizon? What is the equation of state of neutron stars? How did black holes grow, evolve, and influence galaxy formation?

**Strong Gravity**

The observational consequences of strong gravity can be seen close to the event horizon, where the extreme effects of General Relativity (GR) are evident in the form of gravitational redshift, light bending, and frame dragging. The spectral signatures needed to determine the physics of the accretion flow into the black hole are only found in X-rays. IXO will allow us to observe orbiting features from the innermost accretion disk where strong gravity effects dominate (Fig. 1-1; Astro2010 White Paper: Spin and Relativistic Phenomena around Black Holes, Brenneman et al.).

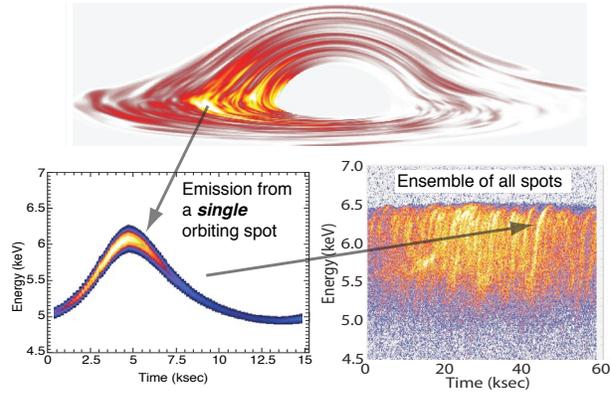

*Figure 1-1. IXO will resolve multiple hot spots in energy and time as they orbit the SMBH, each of which traces the Kerr metric at a particular radius. In the time-energy plane, the emission from these hot spots appears as "arcs," each corresponding to an orbit of a given bright region.*[6]

Observations of SMBH with XMM-Newton have revealed evidence of "hot spots" on the disk that light up in the iron Kα line, allowing us to infer their motions.[5] Each parcel of gas follows a nearly circular orbit around a black hole. Tracing these on sub-orbital timescales, however, requires the large 0.65 m$^2$ effective area around 6 keV provided by IXO. The emission from these hot spots appears as "arcs" in the time-energy plane. GR makes specific predictions for the form of these arcs, and the ensemble of arcs reveals the mass and spin of the black hole and the inclination of the accretion disk. Deviations from the GR predictions will create apparent changes in these parameters as a function of time or hot spot radius. IXO will enable the first orbitally time-resolved studies of 10–20 SMBH and provide a direct probe of the physics of strong gravity.

**Neutron Star Equation of State**

Neutron stars have the highest known matter densities in nature, utterly beyond the densities produced in terrestrial laboratories. The appearance of exotic excitations and phase transitions to strange matter have been predicted, but these predictions are uncertain due to the complexity of Quantum Chromodynamics (QCD) in this high-density regime. These uncertainties lead to widely differing equations of state, each of which imply a different neutron star radius for a given mass.[7] IXO will determine the mass-radius relationship for dozens of neutron stars of various masses with four distinct methods: (1) the gravitational redshift and (2) Doppler shift and broadening of atmospheric absorption lines, (3) pulse timing distortions due to gravitational lensing, and (4) pressure broadening of line profiles, all enabled by high resolution spectroscopy and energy-resolved fast timing (The Behavior of Matter Under Extreme Conditions, Paerels et al.).[8,9] IXO measurements of mass and radius for neutron stars in low-mass X-ray binaries (LMXB) drive the high time resolution requirement, and will distinguish among the allowed models (Fig. 1-2).

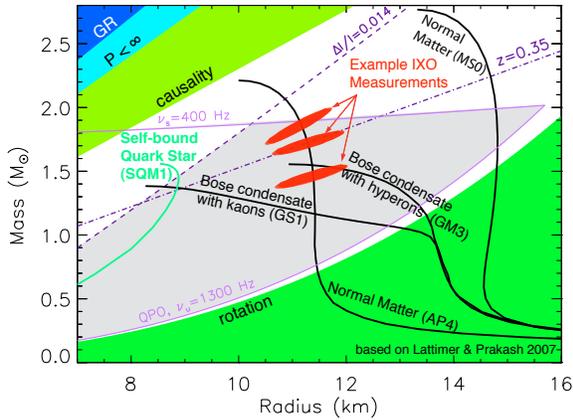

*Figure 1-2. IXO pulse timing measurements (red) of three different LMXBs out of the dozens planned, showing mass and radius values that distinguish among models in the allowed (gray) region.*

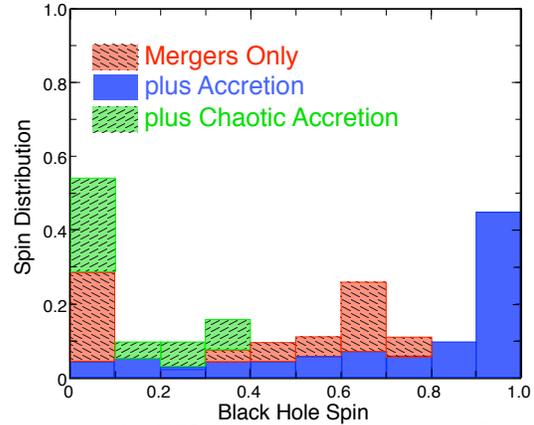

*Figure 1-3. An IXO survey will distinguish between possible spin distributions of z<1 SMBHs resulting from different SMBH evolution models.[13]*

The X-ray polarimeter on IXO will make observations of magnetars that also test predictions of quantum electrodynamics (QED). Magnetars have magnetic fields $B > 4.4 \times 10^{13}$ G where QED predicts novel effects, such as vacuum birefringence. In the presence of matter, resonant polarization mode conversion will occur that will be observed with IXO.[10]

### *Black Hole Evolution*

SMBHs are a critical component in the formation and evolution of galaxies. Future observatories including JWST, ALMA and 30m-class ground-based telescopes will observe the starlight from galaxies out to the highest redshift. IXO will play a crucial role by detecting the accretion power from their embedded SMBHs ($10^7$–$10^9 M_\odot$), *even when obscured*. Luminous, ~$10^9 M_\odot$ SMBHs have been detected at z~6. Growing such massive SMBHs within the <1 Gyr requires sustained Eddington-limited accretion. Gas dynamical simulations predict a period of intense star formation and obscured accretion during the formation of these first galaxies, driven by a rapid sequence of mergers.[11] IXO observations offer the most direct means to discover and study accretion in these systems.

IXO will chart "The Growth of Supermassive Black Holes Over Cosmic Time" (Nandra et al.). Finding growing SMBH at z > 7, which are rare objects, requires a combination of large effective area (3 m$^2$ at 1 keV), good angular resolution (5 arcsec) and large field of view (18 arcmin). These capabilities allow IXO to reach Chandra's limiting sensitivity 20 times faster (Fig. 1-4), enabling the first full characterization of the population of accreting SMBHs at z ~ 7, and constraints at z = 8–10, deep into the cosmic "dark age."

At lower redshifts, z = 1–3, where the majority of accretion and star formation in the Universe occurs, IXO's high throughput for imaging and spectroscopy will uncover and characterize the properties of the most obscured AGN, observing ~10,000 AGN in a 1 Msec 1 sq. deg survey.[12] IXO will also observe the key "blowout" phase in the evolution of massive galaxies, where it is thought that AGN winds expel gas, terminating star formation. This cosmic feedback from SMBHs is a critical ingredient in models of galaxy evolution, which IXO will definitively test by revealing velocity, column density, metallicity, and ionization of the outflows.

Another approach to constraining SMBH evolution is via measurements of their spin.[13] IXO will measure the black hole spin in six independent ways: relativistic disk line spectroscopy, reverberation mapping, disk hot spot mapping, disk continuum spectroscopy, disk polarimetry, and power spectral analysis (Spin and Relativistic Phenomena around Black Holes, Brenneman et al.). A key observational signature is the iron

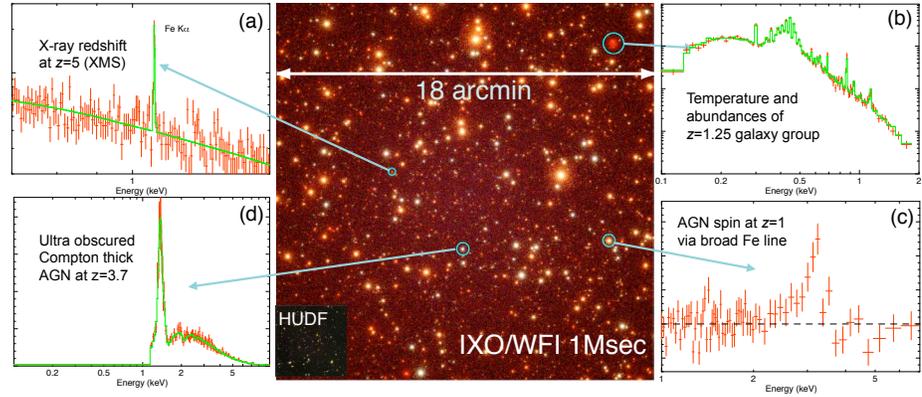

*Figure 1-4. WFI Simulation of the Chandra Deep Field South with Hubble Ultra Deep Field (HUDF) in inset. Simulated spectra of various sources are shown, illustrating IXO's ability (clockwise from top left) to: a) determine redshift autonomously in the X-ray band, b) determine temperatures and abundances even for low luminosity groups to z>1, c) make spin measurements of AGN to a similar redshift, and d) uncover the most heavily obscured, Compton-thick AGN.*

Kα emission line, produced via the illumination of the disk by the primary X-ray continuum and distorted in energy and strength by the gravitational field and relativistic motions around the black hole.

In SMBHs, the spin can be changed by either accretion or merger. The current spin distribution is a record of the relative importance of mergers versus accretion in the growth history of black holes. IXO studies will determine the spin of ~300 SMBHs, sufficient to distinguish merger from accretion models and providing a new constraint on galaxy evolution (Fig. 1-3).

In contrast, the spin of stellar-mass black holes is set at birth, so IXO measurements of ~100 such sources will reveal the angular momentum of the progenitor objects (Stellar-Mass Black Holes and Their Progenitors, Miller et al.).

### *Large Scale Structure*

The extraordinary capabilities of IXO will reveal the major baryonic component of the Universe, in clusters, groups and the intergalactic medium (IGM), and the interplay between these hot baryons and the energetic processes responsible for cosmic feedback. IXO will open a new era in the study of galaxy clusters by directly mapping the gas bulk velocity field and turbulence (Fig. 1-5). IXO's sensitivity will enable us to confront key questions: How does Cosmic Feedback Work? How did Large Scale Structure Evolve? What is the nature of Dark Energy? Where are the Missing Baryons in the Universe?

**Cosmic Feedback from SMBHs**

Energetic processes around black holes result in huge radiative and mechanical outputs (Fundamental Accretion and Ejection Astrophysics, Miller et al.),[14] which can potentially have a profound effect on their larger scale environment in galaxies, clusters and the intergalactic medium (Cosmic Feedback from Massive Black Holes, Fabian et al.). The black hole can heat surrounding gas via its radiative output, and drive outflows via radiation pressure. Mechanical power emerging in winds or jets can also provide heating and pressure. The high spectral resolution and imaging of IXO will provide the necessary spectral diagnostics to distinguish between them.

For outflows that are radiatively accelerated in AGN, X-ray observations will determine the total column density and flow velocity, and hence the kinetic energy flux. IXO will be sensitive to ionization states from Fe I to Fe XXVI over a wide redshift range, allowing the first determination of how feedback affects all phases of interstellar and intergalactic gas, from million-degree collision-ionized plasmas to ten-thousand degree photo-ionized clouds. These measurements will probe over 10 decades in radial scale, from the inner accretion flow where the outflows are generated, to the halos of galaxies and clusters where the outflows deposit their energy.

In the centers of many galaxy clusters, the radiative cooling time of the X-ray-emitting gas is much shorter than the age of the system. Despite this, the gas there is still hot. Mechanical power from the central AGN acting through jets is thought to somehow compensate for the energy lost across scales of tens to hundreds of kpc. IXO will map the gas velocity across dozens of galaxy clusters to an accuracy of tens of km/s, revealing how the mechanical energy is spread and dissipated.

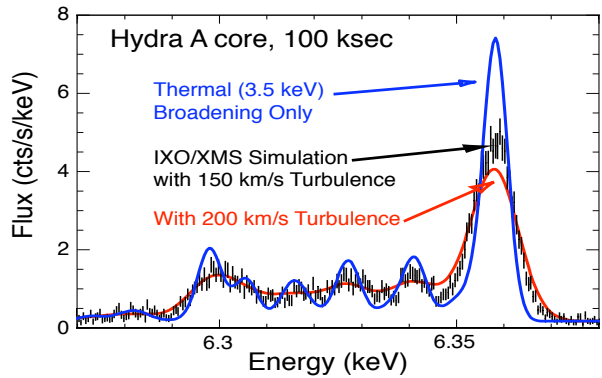

*Figure 1-5. IXO spectrum of Fe XXV lines shows that turbulence of ~ 150 km/s or ~200 km/s may be distinguished from thermal broadening alone. This is currently impossible at CCD resolution. Simulated IXO XMS data in black, models in color.*

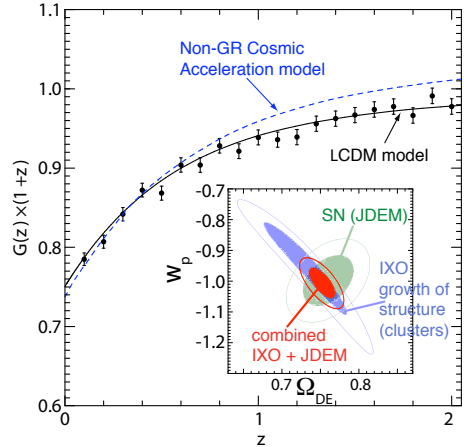

*Figure 1-6. The growth of structure as a function of redshift for simulated observations with IXO of 2000 clusters at z=0–2. The dashed line shows the growth expected in a non-GR models of cosmic acceleration.[18] The inset shows improvements in the constraints on the dark energy equation of state from combination of distance-redshift[19] and structure growth methods.*

**Galaxy Cluster Evolution**

Structure formation is a multi-scale problem. Galaxy formation depends on the physical and chemical properties of the intergalactic medium (IGM). The IGM in turn is affected by energy and metal outflows from galaxies. Detailed studies of the IGM in galaxy clusters are now limited to the relatively nearby Universe (z < 0.5). IXO will measure the thermodynamic properties and metal content of the first low-mass clusters emerging at z ~ 2 and directly trace their evolution into today's massive clusters (Evolution of Galaxy Clusters Across Cosmic Time, Arnaud et al.).

Entropy evolution from the formation epoch onwards is the key to disentangling the various non-gravitational processes: cooling and heating via SMBH feedback and supernova-driven galactic winds. IXO will measure the gas entropy and metallicity of clusters to z ~ 2 to reveal whether the excess energy observed in present-day clusters was introduced early in the formation of the first halos or gradually over time, crucial input to our understanding of galaxy and star formation.

Measuring the evolution of the metal content and abundance pattern of the IGM with IXO will show when and how the metals are produced, in particular the relative contribution of Type Ia and core-collapse supernovae, and the stellar sources of carbon and nitrogen. Precise abundance profiles from IXO measurements will constrain how the metals produced in the galaxies are ejected and redistributed into the intra-cluster medium.

**Cosmology**

The mystery of Dark Energy can be studied by observing the expansion history of the Universe and the growth of matter density perturbations. X-ray observations of galaxy clusters with IXO will provide both tests, complementing other planned cosmological experiments (Cosmological Studies with a Large-Area X-ray Telescope, Vikhlinin et al.). Combining the distance-redshift relation [$d(z)$] and growth of structure data will dramatically improve constraints on the Dark Energy equation of state (Fig. 1-6). These IXO data also test whether the cosmic acceleration is caused by modifications to Einstein's theory of gravity on large scales.

Galaxy cluster observations also provide their own $d(z)$ test by assuming that the mass fraction of hot intracluster gas is constant with redshift. IXO will provide the precise temperature measurements essential to determine the cluster masses. IXO observations of 500 relaxed clusters will give an independent $d(z)$ measurement.[15] The spectral capabilities of IXO will provide direct checks on the relaxed state of the cluster through velocity measurements of the intra-cluster medium.

A recent advance in using galaxy clusters for cosmology was made by combining Chandra observations with advances in numerical modeling, leading to new dark energy constraints from both geometric and growth of structure methods.[16,17] Similarly, combining weak lensing and IXO observations of high-z clusters will reduce systematic errors sufficient to constrain the growth factor to 1% accuracy throughout z = 0–2, leading to very competitive uncertainties in cluster-based cosmological measurements.

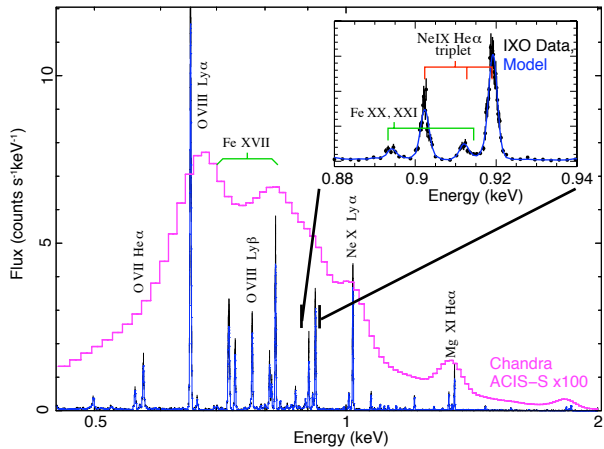

*Figure 1-7. IXO high-resolution X-ray spectra (blue) show the metal-enriched hot gas outflowing from a starburst galaxy, a part of the feedback process unresolvable with current X-ray CCD data (magenta).*

**The Cosmic Web of Baryons**

Less than 10% of the baryons in the local Universe lie in galaxies as stars or cold gas, with the remainder predicted to exist as a dilute gaseous filamentary network—the cosmic web. Some of this cosmic web is detected in Ly$\alpha$ and OVI absorption lines, but half remains undetected. Growth of structure simulations predict that these "missing" baryons are shock heated up to temperatures of $10^7$ K in unvirialized cosmic filaments and chemically enriched by galactic superwinds.[20]

Despite local success in finding hot gas in the halo of the Milky Way, observations with the grating spectrometers on XMM-Newton and Chandra have not yielded conclusive proof for the existence of the hot cosmic web at $z > 0$.[21] The order of magnitude increase in collecting area and R = 3000 spectral resolution of IXO is required to enable detection of the missing baryons and characterize their velocity distribution along at least 30 lines of sight (The Cosmic Web of Baryons, Bregman et al.). This distribution of mass as a function of temperature can be determined from X-ray absorption line grating spectroscopy of highly ionized C, N, and O detected against background AGNs. The extent and nature of galactic superwinds that enrich the web will also be measured both from the proximity of absorption sites to galaxies and the dynamics of the hot gas.

Most galaxies, in fact, have lost more than 2/3 of their baryons, relative to the cosmological ratio of baryons to dark matter.[22] These missing baryons are probably hot, but we do not know if they were expelled as part of a starburst-phase galactic wind, or pre-heated so that they simply never coalesced. X-ray absorption line observations with IXO will, for the first time, identify the location and metallicity of these Local Group baryons from the line centroids and equivalent widths of hot C, N, and O ions, revealing a crucial aspect of galaxy formation (The Missing Baryons in the Milky Way and Local Group, Bregman et al.).[23]

*Life Cycles of Matter and Energy*

The dispersal of metals from galaxies can occur as starbursts drive out hot gas that is both heated and enriched by supernovae. This metal-enriched gas is detected with current X-ray missions, but IXO is needed to measure the hot gas flow velocity using high-throughput spectroscopic imaging (Fig. 1-7), and in turn determine the galactic wind properties and their effects (Starburst Galaxies: Outflows of Metals and Energy into the IGM, Strickland et al.).

The distribution of metal abundances in the Milky Way, including both the gas and dust components, will be mapped using absorption line measurements along hundreds of lines of sight (Measuring the Gas and Dust Composition of the Galactic ISM and Beyond, Lee et al.). On smaller scales, emission from gaseous remnants of supernovae seen with IXO will offer a comprehensive three-dimensional view of the ejecta composition and velocity structure, allowing detailed studies of nucleosynthesis models for individual explosions (Formation of the Elements, Hughes et al.).

IXO will reveal the influence of stars on their local environment via measurements of their coronal activity and stellar winds (Mass-Loss and Magnetic Fields as Revealed Through Stellar X-ray Spectroscopy, Osten et al.). This influence also includes their effect on habitable zones as well as on planet formation. Observations of star-forming regions have shown that X-rays from stellar flares irradiate protoplanetary disks, changing the ion-molecular chemistry as well as inducing disk turbulence.[24] While Chandra has detected a few immense flares,[25] the most significant impact on the protoplanetary disk is in the integrated output of the smaller flares, which can only be characterized using IXO (X-ray Studies of Planetary Systems, Feigelson et al.).

**The full reference list for this document is available at http://ixo.gsfc.nasa.gov/decadal_references/.**

## 2. TECHNOLOGY OVERVIEW

### Observatory Overview

The International X-ray Observatory (IXO) is a facility-class observatory that will be placed via direct insertion into an 800,000 km semi-major axis halo orbit around the Sun-Earth L2 libration point using either an Evolved Expendable Launch Vehicle (EELV) or Ariane V launch vehicle. IXO is built around a large area grazing-incidence mirror assembly with a 20 m focal length. Flight-proven extending masts allow the observatory to fit into either launch vehicle fairing. The mission design life is five years, with consumables sized for 10 years. The observatory design is well defined, building on studies performed over the last decade by NASA, ESA, and JAXA, and has strong heritage from previous space flight missions.

Essential performance parameters derived from the science are shown in Table 2-1. NASA and ESA have each developed a detailed observatory concept that provides these key capabilities. The reference instrument complement, described below, is the same for both concepts while some details of the spacecraft design differ.

### *Payload Overview*

The IXO payload (Fig. 2-1) consists of 1) the Flight Mirror Assembly (FMA), a large area grazing incidence mirror; 2) four instruments mounted in the mirror focal plane on a movable instrument platform (MIP), which are placed in the mirror focus one at a time; and 3) an X-ray Grating Spectrometer (XGS) that intercepts and disperses a fraction of the beam from the mirror onto a CCD camera, operating simultaneously with the observing MIP instrument.

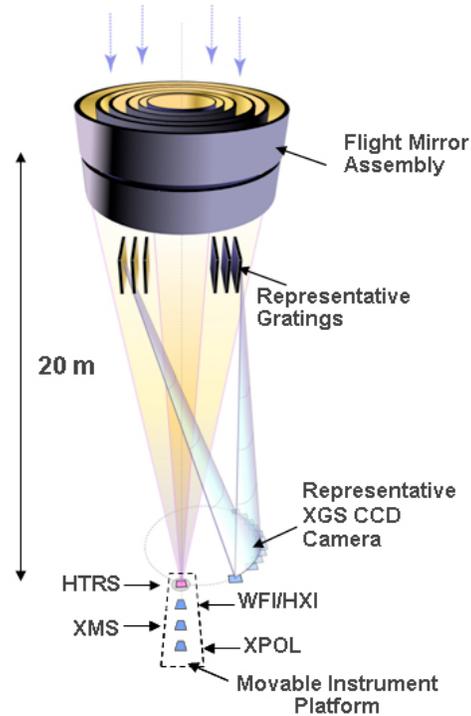

*Figure 2-1. IXO Payload Schematic.*

**The Flight Mirror Assembly** provides effective area of 3 m² at 1.25 keV, 0.65 m² at 6 keV, and 150 cm² at 30 keV. To meet the 5 arcsec mission-level half-power diameter (HPD) requirement for the observatory, the FMA angular resolution must be 4 arcsec or better. Attaining the large effective area within the launch vehicle mass constraint requires a mirror with a high area-to-mass ratio: 20 cm²/kg, 50 times larger than Chandra and eight times larger than XMM-Newton.

As the mirror is the major technical challenge for IXO, two technologies are being developed in a coordinated fashion by NASA, ESA and JAXA as a risk reduction strategy. These are thermally formed ("slumped") glass[26] and silicon pore optics.[27] Both approaches lead to a highly modular mirror design, where the key technology hurdle is the construction of a module. The observatory can accommodate either mirror approach. Both technologies have demonstrated X-ray performance of ~15 arcsec HPD. Details of the ongoing work and plans to achieve the required 4 arcsec HPD (TRL6) are given in Section 3.

Table 2-1. Essential IXO Performance Parameters

| Parameter | Value | | | Science Driver | Inst. |
|---|---|---|---|---|---|
| Mirror Effective Area | 3 m²  @ 1.25 keV<br>0.65 m² @ 6 keV<br>150 cm² @ 30 keV | | | Black Hole Evolution<br>Strong Gravity<br>Strong Gravity | |
| Spectral Resolution (FWHM), FOV, bandpass | $\Delta E$ = 2.5 eV<br>$\Delta E$ = 10 eV<br>$\Delta E$ = 150 eV<br>$E/\Delta E$ = 3000 | 2 arcmin<br>5 arcmin<br>18 arcmin<br>point src | 0.3– 7 keV<br>0.3–10 keV<br>0.1–15 keV<br>0.3– 1 keV | Galaxy Cluster Evolution<br>Cosmic Feedback<br>Black Hole Evolution<br>Cosmic Web | XMS<br>XMS<br>WFI/HXI<br>XGS |
| Angular Resolution | 5 arcsec HPD<br>5 arcsec HPD<br>30 arcsec HPD | | 0.3– 7 keV<br>0.1– 7 keV<br>7–40 keV | Cosmic Feedback<br>Black Hole Evolution<br>Strong Gravity | XMS<br>WFI/HXI<br>WFI/HXI |
| Count Rate | 10⁶ cps with <10% deadtime | | | Neutron Star Eq. of State | HTRS |
| Polarimetry | 1% MDP, 100 ksec,5 × 10⁻¹² cgs (2-6 keV) | | | Strong Gravity | XPOL |

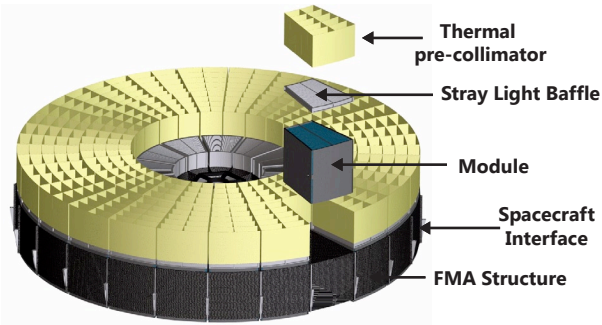

*Figure 2-2. Slumped glass FMA concept. The FMA diameter is 3.3 m.*

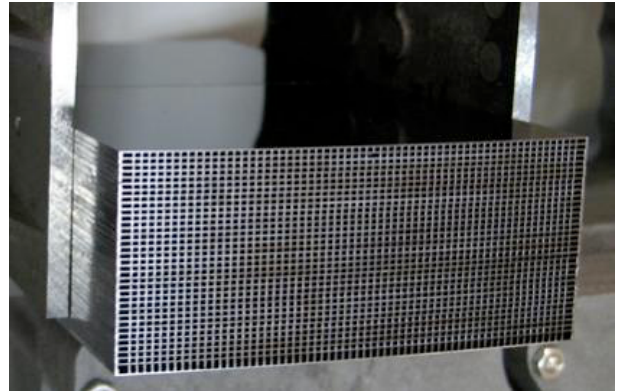

*Figure 2-3. Silicon Pore Optics Module. The module is 6 cm wide.*

The slumped glass FMA design incorporates 361 nested pairs of concentric shells grouped into 60 modules arranged in three concentric rings (Fig. 2-2) with an outer diameter of 3.3 m. Each module comprises approximately 120 pairs of mirror segments, each 20 cm in axial length and 20–40 cm in azimuthal span, accurately aligned in a mounting structure. Segments are produced by thermally slumping 0.4 mm thick glass (same as that manufactured for flat panel displays) onto figured fused quartz mandrels. IXO requires ~14,000 segments. Segment mass production is being demonstrated by NuSTAR for which ~8000 segments are required and the current weekly production rate is ~200. The response above 10 keV is provided by a hard X-ray mirror module mounted in the center of the FMA with segments coated with multilayers to enhance the 10–40 keV reflectivity.

The silicon pore optics approach uses commercial, high-quality 1 mm thick silicon wafers as its base material. One side of a 6-cm-wide rectangular segment of a wafer is structured via etching with accurately wedged ribs approximately 1 mm apart. The other is coated with an X-ray reflecting metallic layer. Segments are then stacked atop an azimuthally curved mandrel and bonded together. This process utilizes techniques and assembly equipment adopted directly from the microelectronics industry. Two stacks are coaligned into a module forming an approximation of a paraboloid-hyperboloid mirror (Fig. 2-3). A total of 236 modules form a "petal," an azimuthal segment of the full mirror. Eight such petals form the complete mirror. The entire production chain—wafer to petal—has been demonstrated. Hard X-ray sensitivity is provided by coating reflecting surfaces at the innermost radii with multilayers.

**The X-ray Microcalorimeter Spectrometer (XMS)** provides high spectral resolution, non-dispersive imaging spectroscopy over a broad energy range. The driving performance requirements are to provide spectral resolution of 2.5 eV over the central 2 × 2 arcmin in the 0.3–7.0 keV band, and 10 eV to the edge of the 5 × 5 arcmin field of view. Its technology status and heritage are discussed in Section 3.

The XMS is composed of an array of microcalorimeters, devices that convert individual incident X-ray photons into heat pulses and measure their energy via precise thermometry. The microcalorimeters are based on Transition-Edge Sensor (TES) thermometers. The extremely rapid change in electrical resistance in the narrow transition (<1 mK) of the superconducting-to-normal transition of the TES allows for extremely accurate thermometry (< 1 μK), thereby enabling < 2.5 eV energy resolution for X-ray photons.

The focal plane consists of a core 40 × 40 array of 300 × 300 μm pixels with spectral resolution of 2.5 eV, corresponding to a 2 × 2 arcmin field of view with 3 arcsec pixels. This core array is surrounded by an outer "annulus" 52 × 52 array of 600 × 600 μm pixels (2,304 pixels total in the outer array) that extends the field of view to 5 × 5 arcmin with better than 10 eV resolution.

The arrays are fabricated using standard micro-electronics techniques. The pixels use Mo/Au bilayer superconducting films deposited on sili-con-nitride membranes in a Si wafer. The X-ray absorbing elements are formed by electroplating Au/Bi films patterned so they provide high array filling factor (95%), but only contact a small area of each TES to prevent electrical and chemical interaction with the sensitive thermometers.

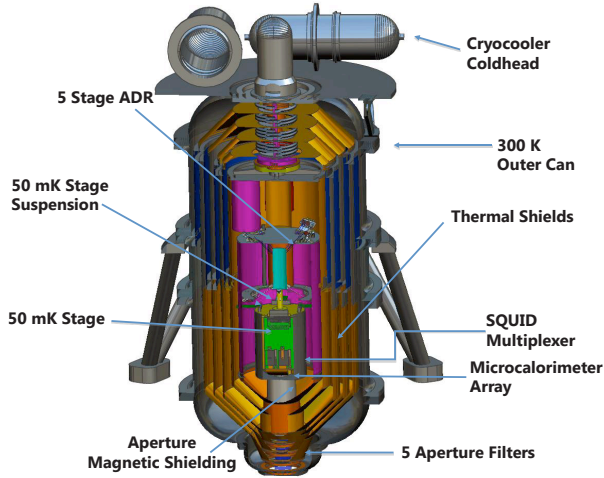

*Figure 2-4. Cutaway view of the XMS dewar assembly (CAD model) including details down to the detector assembly level. The cooler height is just over 1 m.*

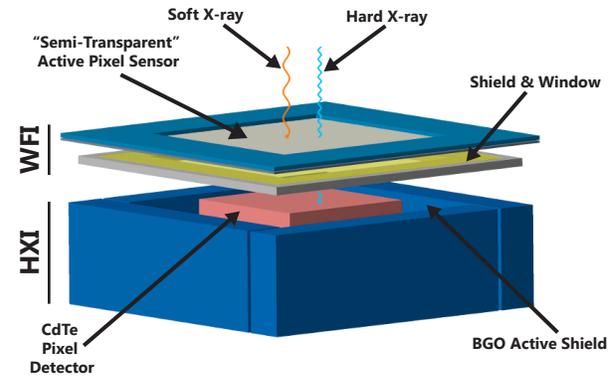

*Figure 2-5. WFI/HXI Layout: The WFI soft X-ray Active Pixel Sensor (APS) is in front of the HXI CdTe detector.*

Currently, 2.3 eV spectral resolution has been demonstrated[28] in a non-multiplexed TES and 2.9 eV has been achieved in a $2 \times 8$ array using a state-of-the-art, time-division SQUID multiplexer system.[29]

A Continuous Adiabatic Demagnetization Refrigerator (CADR) and a mechanical cryocooler provide cooling to 50 mK without expendable cryogens. Figure 2-4 shows a CAD drawing of the XMS cooler assembly.

**The Wide Field & Hard X-ray Imager (WFI/HXI).** The WFI and HXI are two detectors incorporated into one instrument, with the HXI mounted directly behind the WFI (Fig. 2-5).[30] The WFI is an imaging X-ray spectrometer with an 18 × 18 arcmin field of view. It provides images and spectra in the 0.1–15 keV band, with nearly Fano-limited energy resolution (50 eV at 300 eV, < 150 eV at 5.9 keV). A 100 μm × 100 μm pixel size, corresponding to a viewing angle of 1 arcsec, oversamples the beam and thus minimizes pulse pile up.

The WFI's key component is the DEPFET (Depleted P-channel Field Effect Transistor) Active Pixel Sensor (APS). In an APS, each pixel has an integrated amplifier. Compared with earlier CCD-type detectors, the APS concept has the significant advantage that the charge produced by an incident X-ray photon is stored in and read directly from each pixel, rather than being transferred through hundreds or thousands of pixels before being read out. This allows on-demand pixel readout, reduces readout noise, and offers radiation hardness against charge transfer inefficiency.

The DEPFET technology is well tested with numerous prototypes developed for several different missions, including MIXS that will fly on Beppi-Columbo (~2013).[31] Prototype DEPFET devices of 64 × 64 pixels have been tested successfully; an energy resolution at 5.9 keV of 126 eV has been demonstrated. Besides final number of pixels and physical size, all relevant sensor parameters have achieved at least the level required for WFI during prototype characterization measurements.

The HXI extends IXO's energy coverage to 40 keV, observing simultaneously with the WFI.[32] The HXI will have energy resolution better than 1 keV (FWHM) at 30 keV and a FOV of 12 × 12 arcmin. The HXI is a 7 × 7 cm wide Double-sided Strip Cadmium Telluride (DS-CdTe) detector. Its 0.5 mm thickness affords nearly 100% detection efficiency up to 40 keV. To suppress background, an active anticoincidence shield surrounds five sides of the imager. In addition, two layers of Double-sided Si Strip Detector (DSSD) are mounted above the CdTe to serve as particle background detectors and detectors of 7–30 keV X-rays. The IXO HXI is an advanced version of the HXI to fly in 2013 on JAXA's ASTRO-H; all major components have been demonstrated in a laboratory environment.

**The X-ray Grating Spectrometer (XGS)** is a wavelength-dispersive spectrometer for high-res-olution spectroscopy, offering spectral resolution (λ/Δλ) of 3000 (FWHM) and effective area of 1000 cm² across the 0.3–1.0 keV band.

The reference concept incorporates arrays of gratings that intercept a portion of the converging FMA beam and disperse the X-rays onto a CCD array. The existence of two viable grating technologies reduces risk. One implementation utilizes Critical Angle Transmission (CAT) gratings and draw heritage from the Chandra High Energy Transmission Grating Spectrometer, but provide substantially higher efficiency.[33] Another approach incorporates reflection gratings. These so-called "off plane" reflection gratings are ruled along the direction of incidence, rather than perpendicular to it as was implemented on XMM-Newton.[34]

The CAT grating principle has been demonstrated on 3mm × 3mm prototypes, with measured diffraction efficiencies of 80–100% of theoretical values. CAT gratings with a support geometry meeting the IXO requirements have been fabricated. Prototype off-plane gratings are currently undergoing efficiency measurements, with an expectation of obtaining > 40% dispersion efficiency (sum of orders) from 0.3 keV to 1.0 keV.

**The High Time Resolution Spectrometer (HTRS)** will perform precise timing measurements of bright X-ray sources.[35] It can observe sources with fluxes of $10^6$ counts per second in the 0.3–10 keV band without performance degradation, while providing moderate spectral resolution (200 eV FWHM at 6 keV). The HTRS is an array of 37 hexagonal Silicon Drift Diodes (SDD), placed out of focus so that the converging beam from the FMA is distributed over the whole array. The key HTRS performance requirement has been demonstrated using existing detectors and standard analog readout electronics. Development of a fully functional prototype is underway.

**The X-ray Polarimeter (XPOL)** is an imaging polarimeter, with polarization sensitivity of 1% for a source with 2–6 keV flux of $5 \times 10^{-12}$ ergs cm$^{-2}$ s$^{-1}$ (1mCrab).[36] XPOL utilizes a fine grid Gas Pixel Detector to image the tracks of photoelectrons produced by incident X-rays and determine the direction of the primary photoelectron, which conveys information about the polarization of the incoming radiation. The key XPOL performance requirement has been met in the laboratory, and a prototype detector has been vibration and thermal vacuum tested.

## Spacecraft Overview

As mentioned, NASA and ESA have each developed a detailed spacecraft concept (Figs. 2-6 and 2-7). Both concepts are compatible with an EELV and Ariane V. Both studies concluded that the IXO spacecraft could be built with technologies that are fully mature today. All subsystems utilize established hardware with substantial flight heritage. Most components are "off-the-shelf."

The description below is based on the NASA design. The IXO spacecraft concept is robust; all IXO resource margins meet or exceed requirements (Table 2-2). Substantial redundancy along with failsafe mechanisms for contingency mode operations assure that no credible single failure will degrade the mission.

The L2 orbit facilitates high observational efficiency and provides a stable thermal environment. The allowed attitude relative to the sun line is 70°–110° (pitch), ±180° (yaw); ±20° (roll). These ranges keep detectors and radiators out of the sun while providing full illumination to the solar arrays throughout the mission. The space-craft pointing requirement is 10 arcsec (3σ), with post-facto aspect reconstruction accuracy of 1 arcsec; integrated modeling shows these accuracies are achievable with > 50% margin. IXO carries out observations by pointing at celestial objects for durations of $10^3$–$10^5$ sec. Since all the detectors are photon counting, longer integrations can be performed by multiple exposures.

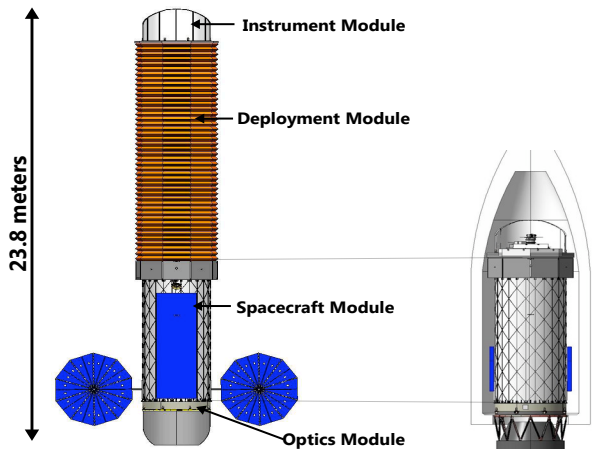

*Figure 2-6. NASA observatory concept, showing on-orbit and launch configurations.*

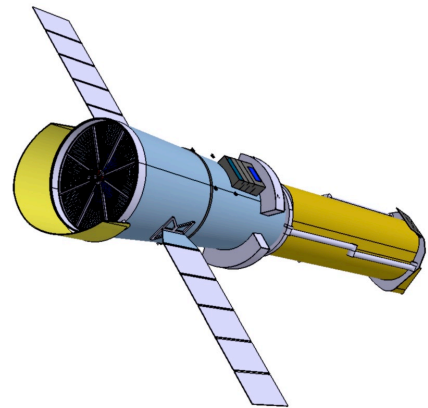

*Figure 2-7. ESA observatory concept.*

Table 2-2. Technical Resources Summary (Mass, Power & Data)

|  | Estimated Value | Growth Contingency % | Maximum Expected Value | Margin % | Minimum Available Resource | Total Growth Reserve % |
|---|---|---|---|---|---|---|
| Launch Mass [kg] | 4779 | 16.5%* | 5570** | 15.3% | 6425 (LV Throw Mass) | 34.4% |
| Power Consumption [W] | 3634 | 30.0% | 4724 | 10.1% | 5200 (S/A Output) | 43.1% |
| Stored Data (24 hr) [Gbits] | 128 | 30.0% | 166 | 141.1% | 400 (On-board Memory) | 213% |

\* Mass Growth Contingency %'s based on AIAA-S-120, Mass Properties Control for Space Systems
\*\* Includes 197 kg (3 σ) propellant for 10 years of operations

IXO consists of four major modules: Instrument, Deployment, Spacecraft, and Optics (Fig. 2-6). This architecture facilitates parallel development and integration and test.

**The Instrument Module (IM)** (Fig. 2-8) accommodates the instruments. All detectors except the XGS camera mount to the movable instrument platform (MIP), which is comparable to moving platforms on Chandra and ROSAT. Focus and translation mechanisms, coupled with a metering structure metrology system based on Chandra heritage, assure centering of the detectors in the converging X-ray beam and accurate attitude reconstruction.

**The Deployment Module (DM)** is the portion of the metering structure which is extended on orbit. It consists of three identical ADAM masts, similar to those on NuSTAR. High precision deployment accuracy/repeatability was proven with the 60 m ADAM used in space on the NASA's Shuttle Radar Topography Mission. As the masts deploy, they pull with them wire harnesses and two pleated shrouds that shield the instruments thermally and from stray light.

**The Spacecraft Module (SM)** accommodates the bulk of the spacecraft subsystems including the power; propulsion; RF communications; guidance, navigation, and control; and avionics. The electronics boxes, reaction wheels, and propulsion tanks mount to a nine-sided honeycomb deck. The 6.6 m × 3.3 m diameter cylindrical composite metering structure accommodates the solar arrays, thrusters, and high-gain antenna.

**The Optics Module (OM)** includes the FMA, its sunshade, and the star trackers. The Optics Module interfaces the FMA to the fixed metering structure within the SM.

## IXO Mission Operations

A reference IXO Operations Concept has been developed that describes the envisioned IXO mission operations approach and architecture of the supporting ground data systems. Primary telemetry, tracking, and command services will be provided by one of the agency's Deep Space Networks (DSN), with short daily ground contacts. Flight and science operations will be conducted from a joint IXO Science and Operations Center (ISOC).

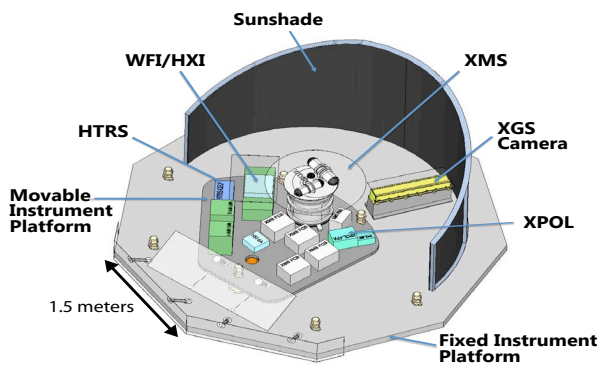

Figure 2-8. Instrument Module with IXO's five instruments, including four on the MIP.

The full reference list for this document is available at http://ixo.gsfc.nasa.gov/decadal_references/ and the acronym list is available at http://ixo.gsfc.nasa.gov/decadal_acronyms/.

# 3. TECHNOLOGY DRIVERS

## IXO X-ray Mirror Technology

The X-ray mirrors are the mission enabling technology for IXO and are also the technology needing the most substantial development. By comparison, all other payload technologies have demonstrated performance at or near requirements and spacecraft technologies are fully mature.

Two approaches for the mirrors are being pursued as a risk reduction strategy: segmented, slumped glass (Fig 3-1) and Si pore optics (Fig 2-4). The IXO Telescope Working Group oversees the progress of the technology development.

Slumped glass mirror segment pairs have been fabricated and aligned that have measured performance of 15 arcsec HPD at 8 keV (Figure 3-2), consistent with modeled performance predictions for the test configuration.[26] Primary sources of blur include the forming mandrel figure, surface roughness in the 0.1–10 mm band imparted during the slumping, and segment fixturing accuracy. Mounted slumped glass mirrors have been successfully vibration tested, and acoustically tested to launch qualification levels.

Silicon pore optics stacks with as many as 45 plates have been assembled using a robotic process. Full X-ray illumination of 4 pairs of plates in an aligned stack pair (a module) yields ~ 17 arcsec HPD.[27] Current performance is limited by particulate contamination present during the stacking process. Initial mechanical pull tests to check integration of the cold welds that hold the stacks together demonstrate the robustness of the silicon pore stacks.

The technology development program is designed to systematically investigate and reduce the sources of blur, and produce modules that both meet the 4 arcsec HPD performance requirement and survive environmental testing. Technology roadmaps for both the silicon pore optics and the slumped glass optics lead to TRL 6 by 2012.

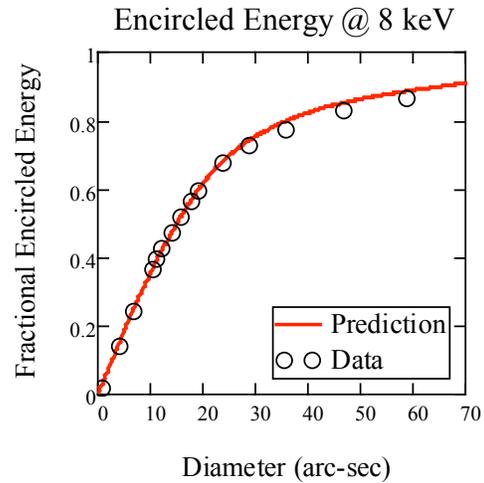

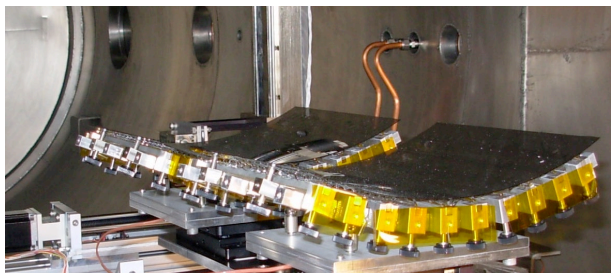

*Figure 3-1. Slumped glass mirror segment pair in the GSFC X-ray testing facility.*

*Figure 3-2. The point spread function measured (circles) at 8 keV for a mounted glass mirror pair is consistent with model predictions (red line).*

The glass roadmap shows separate TRL 5 criteria for segment fabrication and alignment and integration. TRL 5 for segment fabrication entails use of 2 arcsec forming mandrels, reduction of coating stress, and reduction of the mandrel's boron nitride release layer roughness. Two arcsecond forming mandrels have been made. The coating stress reduction has been demonstrated but not yet implemented on segments and the boron nitride work is continuing. For alignment, TRL 5 will be demonstrated by bonding multiple mirror segment pairs into a medium fidelity housing, demonstrating 10 arcsec performance in X-rays, and meeting environmental requirements. TRL 6 entails aligning and bonding multiple pairs of mirror segments into a flight-like housing attaining 4 arcsec performance in X-rays, and meeting environmental requirements.

For pore mirrors, production of a small stack meeting the angular resolution requirement will demonstrate TRL 5. Substantial progress toward this goal is anticipated from systematic particulate removal from plates, and plate assembly in a cleaner environment. TRL 6 entails production of a full module that meets performance and environmental requirements.

The selection of the flight technology will be made at the multi-agency level. Selection will be based on performance, including angular resolution, effective area, mass, cost, and programmatic considerations. Independent of which approach is selected, mirror development will involve participants from the US, Europe, and Japan.

## X-Ray Microcalorimeter Spectrometer

To meet its performance requirements (Table 2-1), the XMS uses a 4-kilo-pixel TES array. The XMS has 40 × 40 pixels in the core array, plus an additional 2,304 pixels in the surrounding outer array. The array will be read out by a multiplexed SQUID amplifier and cooled to 50 mK using a multistage CADR and a expendable-free cryocooler.

TES arrays have been produced with pixels that meet the spectral resolution requirements (Fig, 3-3). The challenge lies in producing full arrays with appropriate readout electronics. 32 × 32 arrays and SQUID multiplexer readouts are being fabricated (Fig. 3-4) to assess energy resolution performance and quantify noise budgets. Work is underway to improve heat sinking so that the energy resolution requirement is met with high uniformity, and to implement multiple absorbers read out by a single TES to extend the field of view.[37]

Two approaches are under development for the readout. A prototype Time-Division-Multiplexing (TDM) SQUID readout system has been successfully tested. Two columns of 8 TESs (2 × 8 demo) were read out, and an average resolution of 2.9 eV was achieved with very high uniformity (± 0.02 eV).[29] An alternate approach, Frequency Division Multiplexing (FDM) with base-band feedback, should be demonstrated later this year.

Multiplexing 32 rows of TES pixels while improving the energy resolution to meet the requirement entails two straightforward improvements to the TDM system: lower SQUID noise and faster switching speed. Reduction in SQUID noise will be achieved by better heat sinking of the multiplexer chip and by implementation of a new generation of quieter series-array SQUIDs placed on the 50 mK stage. Increasing the row-switching speed by a factor of four will allow 32 rows to be read out, with each pixel being sampled at the same rate as currently used to sample eight rows. These advances will be incorporated into a read-out system capable of multiplexing three columns of 32 pixels. Environmental tests will be performed by the end of 2009, bringing the technology to TRL 5. A 6 × 32 readout of both the core and extended arrays will be accomplished by the mid-2013 (TRL 6). The system will be verified at counting rates up to 200 cps per pixel.

The XMS requires a mechanical cryocooler to cool to below 5 K and a CADR to cool from below 5K to ~35 mK. A five-stage CADR is baselined. Flight-qualified single stage ADRs have operated at temperatures as low as 35 mK. A full-scale four-stage CADR breadboard has demonstrated cooling to 50 mK with a 5 K heat sink. The next steps are to qualify a two-stage ADR for the ASTRO-H mission (2013 launch) by mid-2011 (TRL 6), and then fabricate and environmentally test a five-stage system.

The multi-stage mechanical cryocooler must provide 30 mW of cooling power < 5K. Pulse-Tube (PT) and Joule-Thomson (JT) cryocooler technologies have reached high levels of maturity and are being implemented on several flight missions. The 6K cyrocooler for the JWST/MIRI instrument is at TRL 6 and consists of a two-stage PT cooler and a $^4$He JT cooler. Plans are in place to modify the JWST/MIRI JT cooler for use with $^3$He to achieve temperatures below 2K. JT coolers will be flown in 2009 on HERSCHEL and the Japanese SMILES instrument ($^4$He, 4K) for the ISS, the latter meets the IXO requirements. Other cryocoolers under development in US, Europe, and Japan, and based on combinations of Stirling, PT, JT technologies, will meet the IXO requirements.

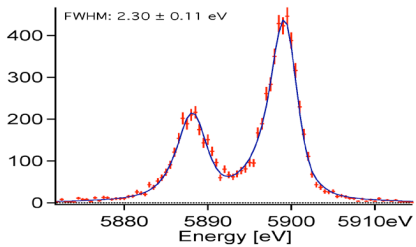

*Figure 3-3. Single-pixel performance from 8x8 X-ray microcalorimeter array with Au/Bi absorbers using $^{55}$Fe X-ray source, demonstrating 2.3 eV resolution (2.9 eV when multiplexed).*

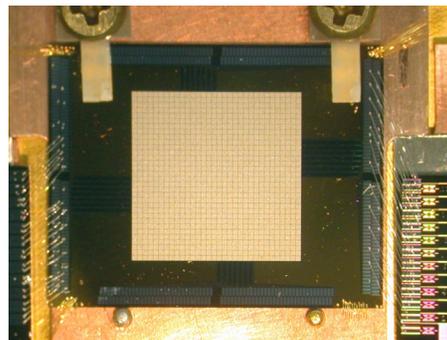

*Figure 3-4. A 32 × 32 microcalorimeter array with 300 micron pixels, high filling factor (93%) and high quantum efficiency (98% at 6 keV).*

## 4. ORGANIZATION, PARTNERSHIP, AND STATUS

The International X-ray Observatory (IXO) mission will be implemented as a tri-agency partnership between the National Aeronautics and Space Administration (NASA), the European Space Agency (ESA), and the Japan Aerospace and Exploration Agency (JAXA). All three agencies have extensive expertise and experience in space science missions, spaceflight system development, and X-ray instrumentation. The agencies have partnered successfully in the past on astrophysics missions such as XMM-Newton (ESA-NASA), ASCA and Suzaku (JAXA-NASA).

IXO is the result of merging the NASA Constellation-X (Con-X) and ESA/JAXA XEUS mission concepts in July 2008. Both mission concepts had been under study by their respective agencies for a decade. Con-X was the second-priority major space initiative for the US in the 2000 National Academy of Science Decadal survey. XEUS was recently selected as one of three candidate large missions in ESA's Cosmic Visions (CV) program.

NASA, ESA, and JAXA established the IXO Study Coordination Group (SCG), via an exchange of letters, to oversee the mission concept development during the current Pre-Phase A stage of the Program. The activities have focused on consolidating the science objectives, mission performance requirements, and observatory design. Three supporting international groups are the Science Definition Team (SDT), the Instrument Working Group (IWG), and Telescope Working Group (TWG). Members of these groups, including the SCG, have substantial experience in developing spaceflight missions and instrumentation in international collaborations.

The SCG oversees NASA and ESA mission concept studies. ESA is currently conducting internal mission-concept studies in preparation for industry studies later in 2009 to support the CV process, for which selections will be made in 2010. NASA is currently conducting internal mission-concept studies, and industry mission studies will be conducted during Phase A.

ESA, NASA, and JAXA will define the mission roles and responsibilities, including the agency providing the launch vehicle, via Letter of Agreement and/or Memorandum of Understanding, by the end of 2012. This is during Phase A and will be based on the output of the mission studies, the mirror technology status, and programmatic considerations. The mirror technology review in mid-2012 will guide the decision regarding the role(s) of the partners in the Flight Mirror Assembly (FMA). The modular nature of IXO provides clearly definable interfaces to enable sharing the development between international partners. We envision the lead agency responsibilities to include Spacecraft Module development, mission systems engineering, and Observatory Integration and Test. The Instrument Module, including its integration with the instruments, can be a separate well defined contribution, as can the Deployment Module.

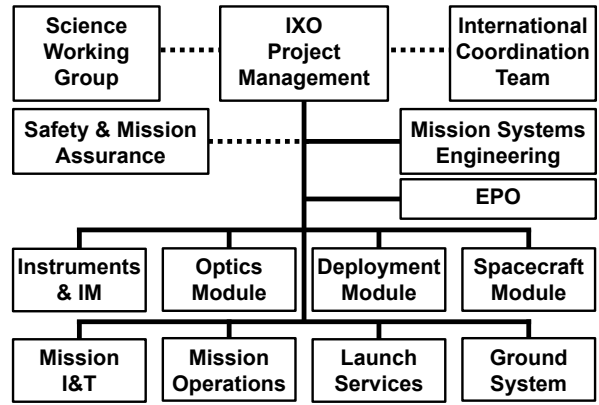

*Figure 4-1. IXO Organization Structure for Mission Development (Phases B - D).*

The instruments will be selected via a coordinated, multi-agency Announcement of Opportunity (AO) process. The AO will include instrument performance specifications commensurate with the capabilities required to achieve the key observatory science objectives. We note that there have been many successful examples of international cooperation in developing X-ray instrumentation, e.g., the gratings on Chandra and XMM-Newton. Within ESA it is typical for instruments to be multi-national consortia and there is heritage for JAXA/NASA collaborations on missions.

A summary-level project organization chart for mission development (Phases B-D) is given in Fig. 4-1. The overall structure reflects the modular approach to the observatory development. Key stakeholders across the IXO team will form the International Coordination Team that will expedite resolution of issues across institutional boundaries. NASA's role in IXO will be managed by GSFC. NASA will competitively select any industry contractors for the spacecraft and/or FMA development.

# 5. SCHEDULE

The top level mission schedule is shown in Fig. 5-1. This schedule supports a May 2021 launch readiness, with eight years and one month for development through on-orbit checkout (Phases B, C, and D). This includes a total of 10 months of funded schedule reserve on the critical path. Five years of mission operations after launch are nominally planned, with an option to extend the science mission to 10 years. Additional schedule reserve is held on the key technology development efforts for the mirror during Phase A and flight development activities not on the critical path.

The schedule reflects our ability to capitalize on the modular nature of IXO. The four Observatory modules (Optics Module, Instrument Module, Spacecraft Module, and Deployment Module) are developed and qualified in parallel, and then delivered for final Observatory Integration and Test (I&T). Milestones and key decision points consistent with NASA 7120.5D are used in the overall planning. Time has been allocated as appropriate for each of the processes for solicitation, selection and contract awards.

The critical path runs through the development of Flight Mirror Assembly (FMA) to Observatory I&T to the launch site operations and launch. A detailed mirror schedule based on the slumped glass approach has been rolled up into the summary schedule. Lessons learned from 1) established NuStar mirror segment facilities and mirror segment fabrication to date, 2) current IXO mirror technology development, and 3) Chandra and Newton XMM mirror calibration and test effort have been accounted for. The overall schedule duration for the FMA using silicon pore optics is expected to be comparable to that for the slumped glass. Observatory I&T reflects activities, flows, and durations that have been developed based on experience from other space observatories of comparable size and type with an emphasis from the Chandra development.

A total of 10 months of schedule reserve has been allocated on the critical path in Phases B through D. Three months schedule reserve is allocated to the FMA development in Phases B/C/D. Six months reserve is held within the Observatory I&T effort of 20 months. One additional month of reserve is held for launch site activities, which has a total of three months and three weeks duration. Schedule reserve is also allocated for activities that are not on the critical path, including the instruments, the Instrument Module, Spacecraft Module, Deployment Module, and Mission Operations and ground system development activities as indicated in Fig. 5-1.

In addition to the schedule reserve on the mission development discussed above, four months of reserve in Phase A are allocated between the mirror TRL 6 demonstrations and the mirror technology review, which takes place in mid-2012. This review initiates the technology selection process which concludes in late 2012 with NASA, ESA, and JAXA finalization of the responsibilities for each agency.

In 2012, Phase A parallel studies of the overall Observatory implementation will be conducted by multiple industry contractors. Following these studies, after the agency roles and responsibilities have been defined, the lead agency will issue a Request for Proposals (RFP) to select the Observatory prime contractor, who will also provide the Spacecraft Module (SM).

The Announcement of Opportunity (AO) for science instrument teams will be released in late 2011 and instrument contracts awarded by the end of 2012. The instrument design reviews (Preliminary Design Review and Critical Design Review) will be timed to support instrument development and will occur prior to similar reviews at the mission level. All instruments will be fully tested and qualified prior to delivery to the Instrument Module (IM). After the instruments are integrated onto the IM, the IM will be tested and delivered for integration with the rest of the observatory in mid-2019. Completion of IM I&T with the instruments prior to final Observatory I&T provides early verification of the instruments in their flight assembly, reducing overall schedule risk.

Spacecraft Module (SM) I&T will begin with delivery of the qualified primary structure and integration of the propulsion system onto the structure. After SM integration and qualification testing, Observatory I&T will commence with integration of the Deployment Module (DM) with the SM. Integration of the IM and the Optics Module (OM) will be next, followed by Observatory environmental and functional testing.

The development of the Ground System (GS), Mission Operations (MO), and science Data Analysis (DA) system will occur in parallel with the observatory development, with major reviews as indicated in Fig. 5-1. Mission level simulations, including the fully integrated observatory and ground systems are also shown. Phase E duration is five years with the option to extend to 10 years.

April 1 Decad
Update 4-1-0
International X-ray Observatory (IXO)**Figure 5-1. IXO Mission Summary Schedule**

16
Section 5 Schedule

International X-ray Observatory (IXO)## ACRONYMS

| | |
|---|---|
| ADAM™ | ABLE Deployable Articulated Mast |
| ADR | Adiabatic Demagnetization Refrigerator |
| AGN | Active Galactic Nucleus |
| ALMA | Atacama Large Millimeter/submillimeter Array |
| AO | Announcement of Opportunity |
| APS | Active Pixel Sensor |
| ASCA | Advanced Satellite for Cosmology and Astrophysics |
| BEPAC | Beyond Einstein Program Assessment Committee |
| BH | Black Hole |
| CAD | Computer-Aided Design |
| CADR | Continuous Adiabatic Demagnetization Refrigerator |
| CAT | Critical Angle Transmission |
| CCD | Charge-Coupled Device |
| CDR | Critical Design Review |
| CESR | Centre d'Etude Spatiale des Rayonnements |
| CfA | Center for Astrophysics |
| ChaMP | Chandra Multiwavelength Project |
| CL | Confidence-level |
| COSMOS | Cosmological Evolution Survey |
| Con-X | Constellation-X |
| COTS | Commercial Off-The-Shelf |
| CV | Cosmic Vision |
| CXC | Chandra X-ray Center |
| DA | Data Analysis |
| DEPFET | Depleted P-channel Field Effect Transistor |
| DS-CdTe | Double-sided Strip Cadmium Telluride |
| DSSD | Double-sided Si Strip Detector |
| EELV | Evolved Expendable Launch Vehicle |
| EOS | Equations of State |
| ESA | European Space Agency |
| FDM | Frequency Division Multiplexing |
| FIP | Fixed Instrument Platform |
| FMA | Flight Mirror Assembly |
| FOV | Field of View |
| FRR | Flight Readiness Review |
| FWHM | Full Width Half Maximum |
| FY | Fiscal Year |
| GR | General Relativity |
| GS | Ground System |
| GSFC | Goddard Space Flight Center |
| HPD | Half Power Diameter |
| HST | Hubble Space Telescope |
| HTRS | High Time Resolution Spectrometer |
| HUDF | Hubble Ultra-Deep Field |
| HXI | Hard X-ray Imager |
| I&T | Integration and Test |
| ICE | Independent Cost Estimate |
| IGM | Intergalactic Medium |
| IM | Instrument Module |
| INAF | Istituto Nazionale di Astrofisica |
| ISAS | Institute of Space and Astronautical Sciences |
| ISOC | IXO Science and Operations Center |
| IWG | Instrument Working Group |
| IXO | International X-ray Observatory |
| IXO SCG | IXO Study Coordination Group |
| JAXA | Japan Aerospace and Exploration Agency |
| JDEM | Joint Dark Energy Mission |
| JKCS | John Keells Computer Services |
| JT | Joule-Thomson |
| JWST | James Webb Space Telescope |
| KDP | Key Decision Point |
| LMXB | Low Mass X-ray Binary |
| LSST | Large Synoptic Survey Telescope |
| MCR | Mission Confirmation Review |
| MDR | Mission Definition Review |
| MEL | Master Equipment List |
| MIP | movable instrument platform |
| MIRI | Mid-Infrared Instrument |
| MIT | Massachusetts Institute of Technology |
| MIXS | Mercury Imaging X-ray Spectrometer |
| MO | Mission Operations |
| MODA | Mission Operations & Data Analysis |
| MPE | Max-Planck-Institut für Extraterrestrische Physik |
| MS | Mission Simulation |
| MUX | Multiplexer |
| NAR | Non-Advocate Review |
| NASA | National Aeronautics and Space Administration |
| NS | Neutron Star |
| NIR | Near-Infrared |
| NuStar | Nuclear Spectroscopic Telescope Array |
| OM | Optics Module |
| PDR | Preliminary Design Review |
| PER | Pre-environmental Review |
| PRICE-H | Parametric Review of Information for Costing and Evaluation Hardware |
| QCD | Quantum Chromodynamics |
| RASS | ROSAT All-Sky Survey |
| RF | Radio Frequency |
| RFP | Request for Proposals |
| RIXOS | ROSAT International X-ray/Optical Survey |
| ROSAT | Röntgensatellit - a German X-ray satellite telescope |
| SAO | Smithsonian Astrophysical Observatory |
| SCG | Study Coordination Group |
| SDD | Silicon Drift Diodes |
| SDT | Science Definition Team |
| SM | Spacecraft Module |
| SMBH | Supermassive Black Holes |
| SMILES | Superconducting Submilimeter-Wave Limb-Emission Sounder |
| SOC | Science Operations Center |
| SQUID | Superconducting Quantum Interference Device |
| TDM | time-division-multiplexing |
| TES | Transition-Edge Sensor |
| TRL | Technology Readiness Level |
| TWG | Telescope Working Group |
| UDS | Ultra-Deep Survey |
| VLA | Very Large Array |
| WBS | Work Breakdown Structure |
| WFI | Wide Field Imager |
| XGS | X-ray Grating Spectrometer |
| XMM | X-ray Multi-Mirror Mission |
| XMS | X-ray Microcalorimeter Spectrometer |
| XPOL | X-ray Polarimeter |
| XRS | X-ray Spectrometer |

**REFERENCES**

1    Silk, J. and Rees, M.J., 1998, AandA, 331, L1
2    Tremaine, S. et al., 2002, ApJ, 574, 740
3    Fabian, A.C. et al., 2003, MNRAS, 344, L433
4    Croton, D. et al., 2006, MNRAS, 365, 11
5    Iwasawa, K., Miniutti, G., and Fabian, A. C., 2004, MNRAS, 355, 1073
6    Armitage, P. and Reynolds, C. S., 2003, MNRAS, 341, 1041
7    Lattimer, J. M., and Prakash, M., 2007, Physics Reports, 442, 109
8    Cottam, J., Paerels, F., and Mendez, M., 2002, Nature, 420, 51
9    Bhattacharyya, S., Miller, M. C., and Lamb, F. K., 2006, ApJ, 664, 1085
10   Lai, D., and Ho, W. C. G., 2003, Phys. Rev. Letters, 91, 071101
11   Li, Y. et al., 2007, ApJ, 665, 187
12   Bauer, F.E. et al., 2004, AJ, 128, 2048
13   Berti, E., and Volonteri, M., 2008, ApJ, 684, 822
14   McNamara, B.R. and Nulsen, P.E.J., 2007, ARAandA, 45, 117
15   Rapetti, D., Allen, S.W., and Mantz, A., 2008, MNRAS, 388, 1265
16   Allen, S.W. et al., 2008, MNRAS, 383, 879
17   Vikhlinin, A. et al., 2009, ApJ, 692, 1060
18   Dvali, G., Gabadadze, G., Porrati, M., 2000, PhysLett B, 485, 208
19   Huterer, D., and Linder, E. V., 2007, PRD, 75, 023519
20   Cen, R., and Fang, T., 2006, ApJ, 650, 573
21   Bregman, J. N., 2007, ARAandA, 45, 221
22   Hoekstra, H. et al., 2005, ApJ, 635, 73
23   Bregman, J. and Lloyd-Davies, E.J., 2007, ApJ, 669, 990
24   Glassgold, A. E., Feigelson, E. D., and Montmerle, T., 2000, in Protostars and Planets IV, 429
25   Imanishi, K, Tsujimoto, M., and Koyama, K., 2001, ApJ, 563, 361
26   Zhang, W. W. et al., 2008, Proc. of SPIE Vol. 7011, 701103
27   Collon, M. J. et al., 2008, Proc. of SPIE Vol. 7011, 70111E
28   Kilbourne, C. A. et al., 2007, Proc. of SPIE Vol. 6686, 668606
29   Kilbourne, C. A. et al., 2008, Proc. of SPIE, Vol. 7011, 701104-701104-12
30   de Korte, P. A.J. et al., 2008, Proc. of SPIE Vol. 7011, 70110A
31   Treis, J., et al., 2008, Proc. of SPIE Vol. 7021, 70210Z
32   Takahashi, T. et al., 2005, Nuclear Instruments and Methods in Physics Research A 541 332–341
33   Heilmann, R. K. et al., 2008, Proc. of SPIE Vol. 7011, 701106
34   McEntaffer, R. L. et al., 2008, Proc. of SPIE Vol. 7011, 701107
35   Barret, D. et al., 2008, Proc. of SPIE Vol. 7011, 70110E
36   Muleri, F. et al., 2008, Nuclear Instruments and Methods in Physics Research A 584 149–159
37   Smith, S. J. et al., 2008, Proceedings of the SPIE, Volume 7011, pp. 701126-701126-8